\documentclass[%
 reprint,
 superscriptaddress,
 amsmath,
 amssymb,
 aps,
 pra,
]{revtex4-2}

\usepackage{graphicx}
\usepackage{dcolumn}
\usepackage{bm}
\usepackage{hyperref}

\newcommand{\ot}{\omega t}
\usepackage{subfigure}
\usepackage{comment}
\usepackage[x11names]{xcolor}
\usepackage{color,soul}

\newcommand{\hly}{} 
\newcommand{\hlym}{} 
\newcommand{\hlys}{} 
\newcommand{\hlb}{} 
\newcommand{\hlg}{} 
\newcommand{\hlgm}{} 

\begin{document}

\title{Topological Floquet engineering of a three-band optical lattice with dual-mode resonant driving}

\author{Dalmin Bae}
\affiliation{Department of Physics and Astronomy, Seoul National University, Seoul 08826, Korea}

\author{Junyoung Park}
\affiliation{Department of Physics and Astronomy, Seoul National University, Seoul 08826, Korea}

\author{Myeonghyeon Kim}
\affiliation{Department of Physics and Astronomy, Seoul National University, Seoul 08826, Korea}

\author{Haneul Kwak}
\affiliation{Department of Physics and Astronomy, Seoul National University, Seoul 08826, Korea}

\author{Junhwan Kwon}
\affiliation{Department of Physics and Astronomy, Seoul National University, Seoul 08826, Korea}

\author{Y. Shin}
\email{yishin@snu.ac.kr}
\affiliation{Department of Physics and Astronomy, Seoul National University, Seoul 08826, Korea}
\affiliation{Institute of Applied Physics, Seoul National University, Seoul 08826, Korea}

\date{\today}

\begin{abstract}
We present a Floquet framework for controlling topological features of a one-dimensional optical lattice system with dual-mode resonant driving, in which both the amplitude and phase of the lattice potential are modulated simultaneously.
We investigate a three-band model consisting of the three lowest orbitals and elucidate the formation of a cross-linked two-leg ladder through an indirect interband coupling via an off-resonant band.
We numerically demonstrate the emergence of topologically nontrivial bands within the driven system, and a topological charge pumping phenomenon with cyclic parameter changes in the dual-mode resonant driving.
Finally, we show that the band topology in the driven three-band system is protected by parity-time reversal symmetry.
\end{abstract}

\maketitle

\section{Introduction}
\label{sec:Introduction}

Ultracold atoms in optical lattices provide a flexible platform to explore topological insulators and associated phenomena, facilitated by the ability to adjust the lattice configuration experimentally~\cite{PhysRevLett.111.185301, Jotzu2014, Goldman2016, RevModPhys.89.011004, RevModPhys.91.015005}. Periodic time-dependent modulation techniques, also known as Floquet engineering, have been established as an effective method to examine topological bands within these systems. Tailored modulations of the lattice have successfully produced nontrivial bands with novel topological characteristics~\cite{PhysRevB.82.235114, PhysRevLett.109.145301, https://doi.org/10.1002/pssr.201206451, PhysRevX.4.031027, PhysRevA.89.061603, doi:10.1080/00018732.2015.1055918, Kang_2020}, which have led to the observation of many interesting phenomena, including topological charge pumping~\cite{PhysRevLett.129.053201, PhysRevA.106.L051301, Citro2023, Walter2023}. Floquet band engineering has thus become a prominent path in the field of optical lattice research.

\begin{figure}[t]
\includegraphics{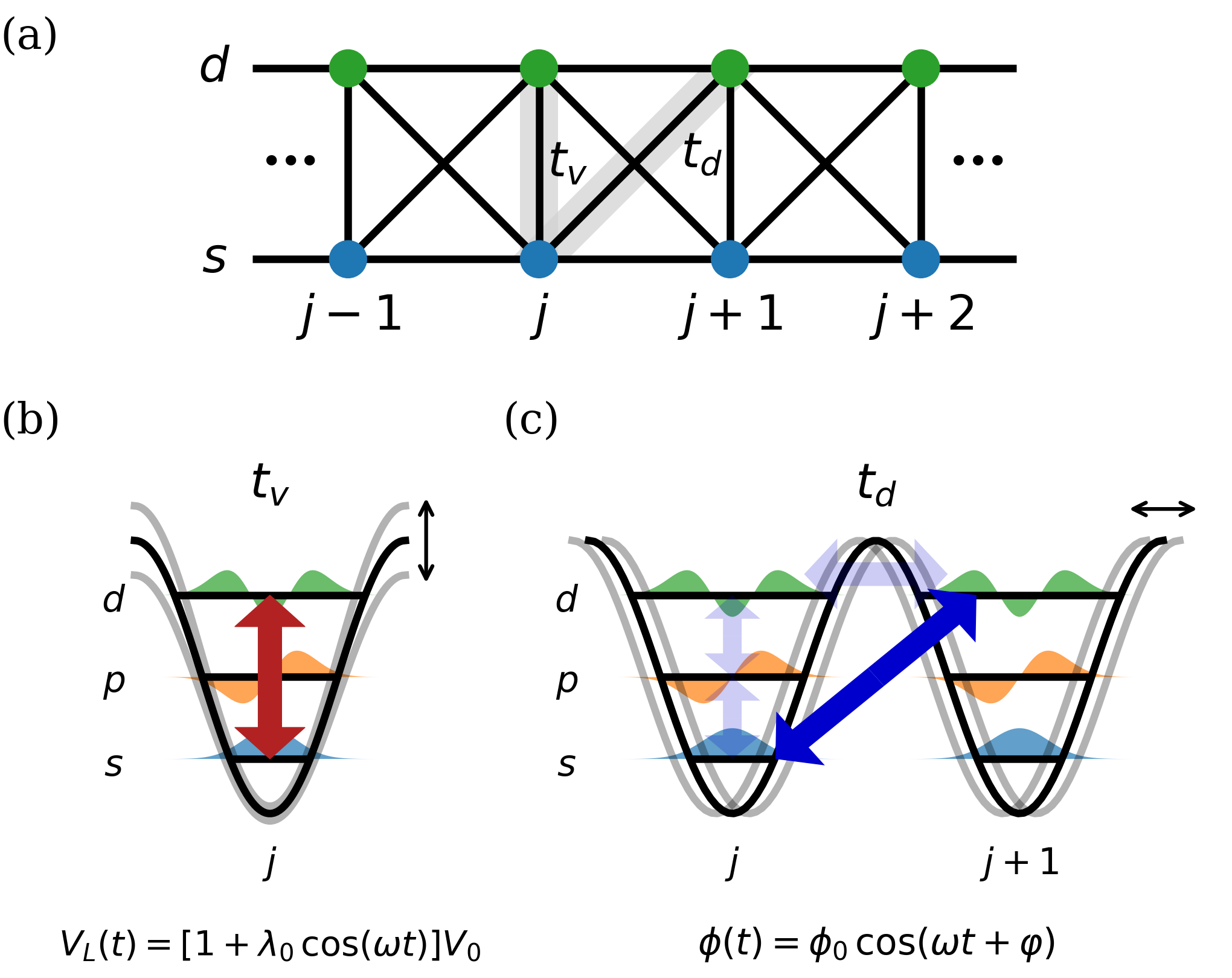}
\caption{\label{fig:intro}(a) Effective ladder model of a 1D optical lattice under dual-mode resonant driving. $s$ and $d$ orbitals comprise the two legs of the ladder, and the vertical ($t_v$) and diagonal ($t_d$) interleg links are formed by (b) the one-photon coupling from the amplitude modulation (AM) of the lattice potential and (c) the three-photon coupling from the phase modulation (PM) that shakes the lattice, respectively. $V_L$ and $\phi$ denote the amplitude and phase of the lattice potential, respectively.}
\end{figure}

Researchers have extensively studied topological bands in one-dimensional (1D) optical lattices to gain essential insight into topological matter. As a minimal representation for 1D topological insulators, in particular, a cross-linked two-leg ladder system or similar models have been investigated~\cite{PhysRevLett.83.2636, PhysRevA.89.023619, PhysRevX.7.031057, Kang_2020, PhysRevResearch.4.013056}. As illustrated in Fig.~\ref{fig:intro}(a), the ladder system is composed of two lines of lattice sites called {\it legs}, and the legs are interconnected both vertically and diagonally, representing the hopping between sites. The diagonal cross-links give rise to topological features in the system. In experimental setups, the legs can be assigned to different spin states of atoms or different orbitals in the lattice, with the cross-linking provided by spin-orbit coupling or band-mixing processes, respectively. In recent experiments, a cross-linked two-leg ladder system employing $s$ and $p$ orbitals was implemented successfully using a two-tone driving scheme~\cite{Kang_2020, PhysRevResearch.4.013056}, where the optical lattices were shaken resonantly with two frequencies, and the cross links were produced by two-photon resonant interband coupling~\cite{PhysRevA.90.051601}. Furthermore, the ability to dynamically adjust the linking properties enabled the demonstration of topological charge pumping~\cite{PhysRevA.102.063315,PhysRevLett.129.053201}.

In this work, we propose an alternative Floquet approach to construct a tunable cross-linked two-leg ladder system. Our approach features creating the ladder with $s$ and $d$ orbitals, which share the same parity, and using both the amplitude and phase modulations of the lattice potential \hlb{simultaneously} at an identical frequency. When the modulation frequency is set close to the energy gap between the $s$ and $d$ bands, the amplitude modulation (AM) generates the on-site resonant coupling between the $s$ and $d$ orbitals, thus forming the ladder rungs [Fig.~\ref{fig:intro}(b)]~\cite{PhysRevA.89.061603, PhysRevA.90.051601, SträterEckardt+2016+909+920, Cabrera-Gutiérrez2019}. Meanwhile, the phase modulation (PM), which triggers lattice shaking, does not generate a direct $s$-$d$ interorbital coupling owing to parity conservation; however, it establishes diagonal connections through three-photon resonant transitions via $p$ orbital [Fig.~\ref{fig:intro}(c)].
This three-photon process represents an {\it indirect} resonant interband coupling that employs an {\it off-resonant} third band as an intermediate state.
To the best of our knowledge, such indirect resonant coupling has not been discussed as an effective interband coupling mechanism in the literature on Floquet band engineering.
Owing to the dual-mode driving employing both AM and PM \hlb{simultaneously}, a cross-linked ladder is formed, comprising two orbitals with identical parity,
\hlg{
which leads to the formation of topological bands that exhibit minimal or absent bulk gaps.
Consequently, this method enables the investigation of the physics of topological semimetals~\mbox{\cite{Burkov2016,Sun2012,Jangjan2021}}, which was not possible in previous studies using lattice shaking.
}

Using a three-band model, we numerically demonstrate the topological properties of the 1D optical lattice system subjected to dual-mode resonant driving.
We comprehensively analyzed the resultant Floquet bands under a range of driving parameter conditions, including the relative intensity and phase of AM and PM. 
Our analysis shows the emergence of a topologically nontrivial phase under certain driving conditions, as evidenced by the entanglement entropy and spectrum~\cite{PhysRevLett.101.010504, PhysRevB.82.241102, PhysRevB.83.245132, PhysRevB.94.205422, PhysRevResearch.4.043164, PhysRevB.73.245115}, along with the observation of a topological phase transition.
Through numerical simulations, we illustrate a topological charge pumping effect expected during slow cyclic changes in driving parameters~\cite{PhysRevB.27.6083, PhysRevLett.111.026802, PhysRevA.90.063638, Nakajima2016, Lohse2016}.
Lastly, we elucidate that the topological phases of the Floquet bands in the three-band model are protected by {\it parity-time reversal} $(PT)$ symmetry.

The remainder of the paper is organized as follows.
Sec.~\ref{sec:2} introduces a three-band model of the 1D optical lattice system under dual-mode resonant driving. We further derive an effective two-band description of the system by adiabatic elimination of the off-resonant $p$ band~\cite{Brion_2007}, which provides insight into the indirect resonant interband coupling and the topological structure of the driven system.
Sec.~\ref{sec:3} presents our numerical results of the quasi-energy and entanglement spectrum of the driven lattice system, and also illustrates the topological charge pumping effect with cyclic parameter changes in the dual-mode resonant driving.
Sec.~\ref{sec:4} demonstrates the role of $PT$ symmetry in protecting the topology of the Floquet bands.
Finally, Section~\ref{sec:Summary} provides a summary and some concluding remarks.

\section{\label{sec:2}Dual-mode resonant driving of optical lattice}
\subsection{\label{subsec:three-band}Three-band model}
Let us consider a spinless fermionic atom in the driven 1D optical lattice potential $V_{\rm lat}(x,t)$, which is given by
\begin{equation}
V_{\rm lat}(x,t) = V_{L}(t) \sin^2\left(\frac{\pi}{a}x - \phi(t)\right),
\end{equation}
where $V_{L}(t)$ and $\phi(t)$ are the amplitude and phase of the lattice potential, respectively, and $a$ is the lattice constant.
$V_{L}$ and $\phi$ are determined by the parameters of the laser beams involved, such as intensity, polarization, and phase, and can be dynamically controlled for Floquet engineering.
The two fundamental modulation approaches are periodically modulating $V_{L}$ and $\phi$ in time, which we refer to as AM and PM, respectively [Figs.~\ref{fig:intro}(b) and \ref{fig:intro}(c)]. As the position of the lattice site is determined by the phase $\phi(t)$, PM induces lattice shaking.
When viewed from the reference frame comoving with the driven optical lattice, the system's Hamiltonian is described as follows~\hly{\mbox{\cite{RevModPhys.91.015005, PhysRevX.4.031027}}}:
\begin{eqnarray}
\label{eq:initial Hamiltonian}
 H(x,t) &=& H_0 + \lambda(t) V_{\rm stat}(x) - F(t) x \\
 H_0 &=& \frac{p^2}{2m} +  V_{\rm stat}(x), \nonumber
\end{eqnarray}
where $p$ is the kinetic momentum of the atom, $m$ denotes its mass, $V_{\rm stat}(x)=V_0\sin^2\left(\frac{\pi}{a}x\right)$ is the stationary lattice potential, $\lambda(t)$ denotes the relative variation of lattice amplitude such that $V_{L}(t)=[1+\lambda(t)]V_0$, and $F(t)=-m\left(\frac{a}{\pi}\ddot{\phi}(t)\right)$ represents the inertial force resulting from PM.

In the tight-binding approximation, the Hamiltonian can be expressed in terms of Wannier states $|j,\alpha\rangle$ localized on lattice site $j$ in the $\alpha$ band, given by~\cite{PhysRevX.4.031027}
\begin{eqnarray}
H(x,t) = &&\sum_{j\alpha}\epsilon_{\alpha}\hat{c}^{\dag}_{j\alpha}\hat{c}_{j\alpha}
-\sum_{jl\alpha}t^{(l)}_{\alpha}e^{-il\theta(t)}\hat{c}^{\dag}_{j\alpha}\hat{c}_{j+l\:\alpha} \nonumber \\
&&+\sum_{jl\alpha\beta}\Big(\lambda(t)u^{(l)}_{\alpha\beta}-F(t)\eta^{(l)}_{\alpha\beta}\Big)e^{-il\theta(t)}\hat{c}^{\dag}_{j\alpha}\hat{c}_{j+l\:\beta}, \nonumber \\
\,
\end{eqnarray}
where $\hat{c}^{\dag}_{j\alpha}\, (\hat{c}_{j\alpha})$ is the creation (annihilation) operator for the atom in the Wannier state $|j,\alpha\rangle$, $\epsilon_\alpha = \langle j,\alpha |H_0| j,\alpha\rangle$ represents the on-site energy, and
\hlg{$t^{(l)}_\alpha = -\langle j,\alpha |H_0| j+l,\alpha\rangle$}
denotes the hopping amplitude between the Wannier states in the $\alpha$ band separated by $l$ lattice sites.
In addition, $u^{(l)}_{\alpha\beta} = \langle j,\alpha|V_{\rm stat}(x)|j+l,\beta\rangle$ and 
\hlg{$\eta^{(l)}_{\alpha\beta} = \langle j,\alpha|x|j+l,\beta\rangle$}
correspond to the lattice potential and lattice displacement matrix elements for interorbital transitions separated by $l$ lattice sites, respectively.
Lastly, $\theta(t)=-\frac{a}{\hbar} \int_{0}^{t} dt' \,F(t')$ represents the time-dependent Peierls phase~\cite{PhysRevA.92.043621}.
See Appendix~\ref{app:Bloch_H} for detailed definitions of the tight-binding parameters.
By Fourier transforming this tight-binding model Hamiltonian, we obtain the Bloch Hamiltonian for quasimomentum $q$ in the presence of AM and PM as follows:
\begin{eqnarray}
H(q,t) = \sum_{\alpha}\Big(\epsilon_{\alpha}-\sum_{l>0}2t^{(l)}_{\alpha}{\cos}[l(q-\theta(t))]\Big)\hat{c}^{\dag}_{q\alpha}\hat{c}_{q\alpha} \nonumber \\
+\sum_{l\alpha\beta}\Big(\lambda(t)u^{(l)}_{\alpha\beta}-F(t)\eta^{(l)}_{\alpha\beta}\Big)e^{il(q-\theta(t))}\hat{c}^{\dag}_{q\alpha}\hat{c}_{q\beta}. \nonumber \\
\,
\end{eqnarray}
Here, $q$ is expressed in units of $1/a$.

In this work, we consider a model system that includes only the three lowest bands, indexed by $\alpha \in \{s, p, d\}$. Considering the lowest-order effects of lattice modulation, the Bloch Hamiltonian of the three-band system is given by 
\begin{equation}
\label{eq:Bloch_Hamiltonian}
H(q,t) =
\begin{pmatrix}
\epsilon_{s}'(q,t) && -F(t)\eta^{(0)}_{sp} && \lambda(t)u^{(0)}_{sd} \\
-F(t)\eta^{(0)}_{ps} && \epsilon_{p}'(q,t) && -F(t)\eta^{(0)}_{pd} \\
\lambda(t)u^{(0)}_{ds} && -F(t)\eta^{(0)}_{dp} && \epsilon_{d}'(q,t)
\end{pmatrix}
\end{equation}
with $\epsilon_{\alpha}'(q,t) = \epsilon_{\alpha}-2t^{(1)}_{\alpha}\cos(q-\theta(t))+\lambda(t)u^{(0)}_{\alpha\alpha}$.
\hly{See Appendix~\mbox{\ref{app:Bloch_H}} for details on the derivation.}

We focus on a case where the system is subjected to dual-mode resonant driving with 
\begin{eqnarray}\label{eq:modulations}
\lambda(t) ~&&= \lambda_0 \cos(\ot), \nonumber \\
\phi(t) ~&&= \phi_0 \cos(\ot+\varphi),
\end{eqnarray}
and the driving frequency $\omega \approx \omega_{sd} = (\epsilon_d - \epsilon_s)/\hbar$.
Here, $\lambda_0$ and $\phi_0$ are dimensionless parameters that represent the strengths of AM and PM, respectively, and $\varphi$ is the relative phase of the two modulation modes.

\subsection{\label{subsec:Effective}Effective two-band model}

\begin{figure}[t]
\includegraphics{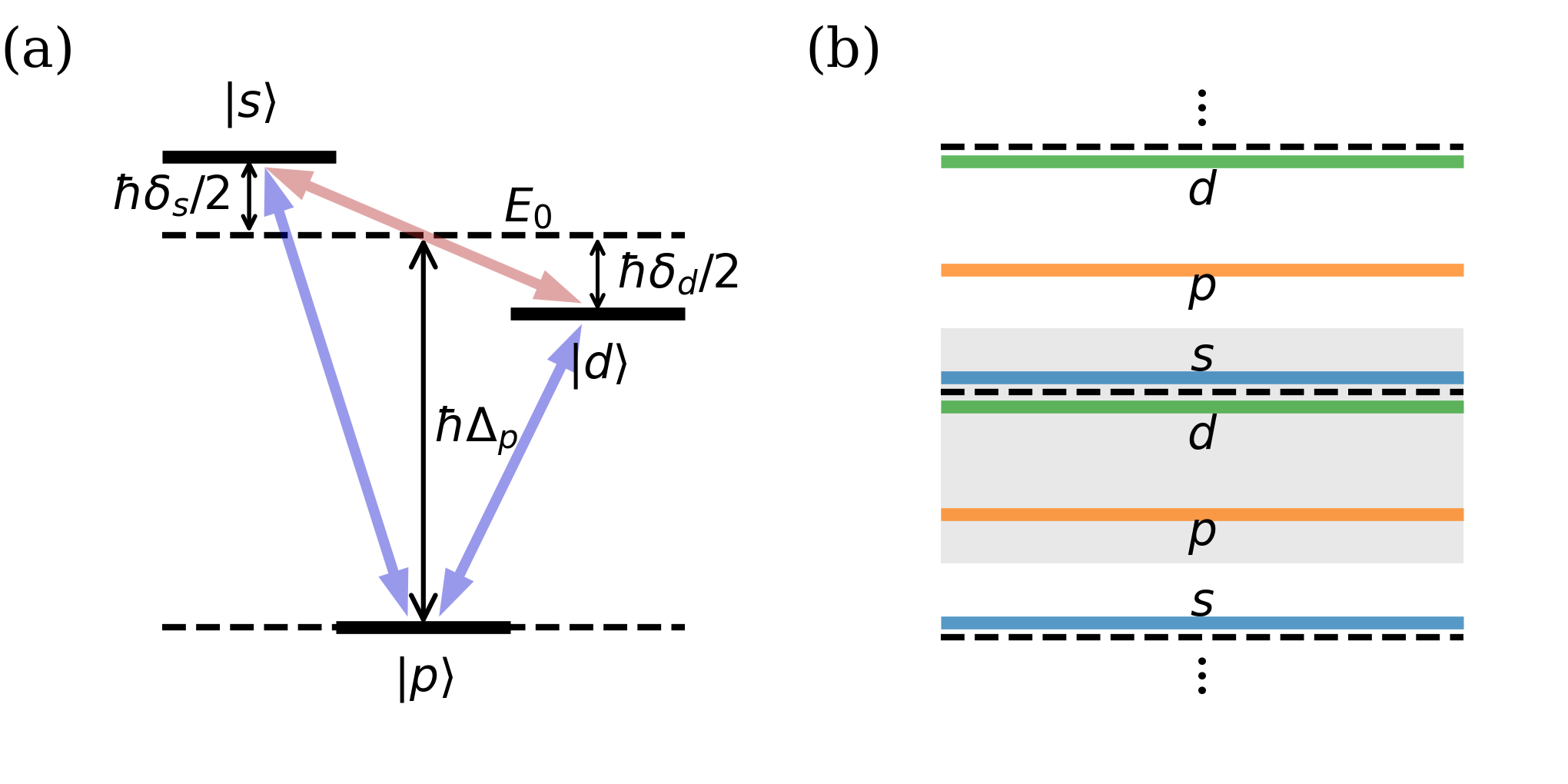}
\caption{\label{fig:V-type}(a) Energy level scheme of the driven three-band system in a rotating frame. \hlb{The red and blue arrows indicate couplings between the two adjacent upper states by $\lambda(t)$ and between an upper state and a lower state by $F(t)$, respectively. $E_0$ denotes the zero energy point.} (b) Floquet energy diagram with a driving frequency $\omega \approx \omega_{sd}$.}
\end{figure}

When the three-band lattice system is driven with a frequency $\omega \approx \omega_{sd}$, the couplings between the $p$ orbital and the others become off-resonant, resulting in the $p$ band being energetically isolated.
We can project the three-band system into an effective two-band system using an adiabatic elimination technique~\cite{Brion_2007} owing to the minimal involvement of the \hlb{$p$} band in band mixing.

\begin{table*}
\caption{\label{tab:parameters}
Tight-binding parameters of optical lattice for $V_0 = 10\,E_R$, where $E_R=\frac{(\hbar k_L)^2}{2m}$ is the recoil energy with $k_L=\pi/a$, and the parameters of the effective two-band model in Eq.~(\ref{eq:eff_H_1}). The values of the effective two-band parameters were calculated for $\lambda_0 = 0.1$, $\phi_0 = 0.1$, and $\omega = \omega_{sd} = (\epsilon_{d}-\epsilon_s)/\hbar$.}
\renewcommand{\arraystretch}{2}
\begin{ruledtabular}
\begin{tabular}{cccccccl}
\multicolumn{5}{c}{Tight-binding parameters} & \multicolumn{3}{c}{Effective two-band parameters}\\
\cline{1-5}
\cline{6-8}
& & $s$ & $p$ & $d$ & $t_{-}$ & $\frac{t^{(1)}_d-t^{(1)}_s}{2}$ & 0.388$\,E_R$ \\ \cline{3-5}
$\epsilon_\alpha$ & $\langle j,\alpha |H_0| j,\alpha \rangle$ & 2.885$\,E_R$ & 7.933$\,E_R$ & 12.059$\,E_R$ & $\lambda'$ & $\lambda_0 u^{(0)}_{sd}$ & 0.173$\,E_R$ \\
$t^{(1)}_\alpha$ & $-\langle j,\alpha |H_0| j+1,\alpha \rangle$ & 0.019$\,E_R$ & 0.244$\,E_R$ & 0.794$\,E_R$ & ${F'}^{2}$ & $\frac{\eta^{(0)}_{sp}\eta^{(0)}_{pd}}{2\hbar\Delta_p}{F_0}^2$ & 0.663$\,E_R$ \\
$u^{(0)}_{\alpha\alpha}$ & $\langle j,\alpha |V_{\rm stat}(x)| j,\alpha \rangle$ & 1.602$\,E_R$ & 4.832$\,E_R$ & 6.315$\,E_R$ & $\lambda'_{-}$ & \hlb{$\lambda_0\frac{u^{(0)}_{dd}-u^{(0)}_{ss}}{2}$} & 0.236$\,E_R$ \\
& & $sp$ & $pd$ & $sd$ & ${F'_-}^{2}$ & $\frac{{\eta^{(0)}_{pd}}^2-{\eta^{(0)}_{sp}}^2}{4\hbar\Delta_p}{F_0}^2$ & 0.317$\,E_R$ \\ \cline{3-5}
$u^{(0)}_{\alpha\beta}$ & $\langle j,\alpha |V_{\rm stat}(x)| j,\beta \rangle$ & 0 & 0 & 1.725$\,E_R$ & $F_0$ & $m{\omega}^2\frac{a}{\pi}\phi_0$ & 4.220$\,E_R k_L$ \\
$\eta^{(0)}_{\alpha\beta}$ & $\langle j,\alpha |x| j,\beta \rangle$ & 0.440$\,/k_L$ & 0.698$\,/k_L$ & 0 & $\theta_0$ & $-\frac{a}{\hbar\omega}F_0$ & -1.443 \\
\end{tabular}
\end{ruledtabular}
\end{table*}

First, let us take a proper rotating frame by applying a unitary transformation of $U_R(t)=\exp{(+i \hat{R} t)}$ to the Bloch Hamiltonian $H(q,t)$ in Eq.~(\ref{eq:Bloch_Hamiltonian}), where
\begin{equation}
\hat{R} = 
{\begin{pmatrix}
-\omega + E_0/\hbar & 0 & 0 \\
0 & E_0/\hbar & 0 \\
0 & 0 & E_0/\hbar
\end{pmatrix}}    
\end{equation}
with $E_0 = (\epsilon_d + \epsilon_s + \hbar\omega)/2$ representing the zero energy point. In the rotating frame, the modified Hamiltonian $H'(q,t)$ is given by
\begin{eqnarray}
H&&'(q,t) \nonumber \\
&&\hlym{=U_{R}(t)H(q,t)U^{\dag}_{R}(t)+i\hbar\dot{U}_{R}(t){U}^{\dag}_{R}(t)} \nonumber \\[5pt]
&&={\begin{pmatrix}
\hbar\delta_s/2 & \hlym{-F(t)\eta^{(0)}_{sp}e^{-i\ot}} & \hlym{\lambda(t)u^{(0)}_{sd}e^{-i\ot}} \\
\hlym{-F(t)\eta^{(0)}_{ps}e^{i\ot}} & -\hbar\Delta_p & -F(t)\eta^{(0)}_{pd} \\
\hlym{\lambda(t)u^{(0)}_{ds}e^{i\ot}} & -F(t)\eta^{(0)}_{dp} & -\hbar\delta_d/2
\end{pmatrix}} \nonumber \\
\end{eqnarray}
with
$\hbar\delta_{s}/2 = \epsilon_{s}' + \hbar\omega - E_0$,
$\hbar\delta_{d}/2 = E_0 - \epsilon_{d}'$, and
$\hbar\Delta_p = E_0 - \epsilon_{p}'$.
Note that $|\Delta_p| \gg |\delta_s|,|\delta_d|$ when the driving frequency is set to $\omega \approx \omega_{sd}$, providing a suitable condition for adiabatic elimination of the $p$ band.
The energy level structure is depicted in Fig.~\ref{fig:V-type}(a). It can be viewed as a characteristic V-type system in which the two adjacent upper states are coupled to each other by $\lambda(t)$ and also to a lower level simultaneously by $F(t)$. For comparison, the Floquet energy diagram of the driven three-band system is illustrated in Fig.~\ref{fig:V-type}(b).

Simplifying the notation of $H'(q,t)$ as
\begin{equation}
H'(q,t) = {\begin{pmatrix}
H_{00} & H_{01} & H_{02} \\
H_{10} & H_{11} & H_{12} \\
H_{20} & H_{21} & H_{22}
\end{pmatrix}},
\end{equation}
the equation of motion for the system state $|\psi\rangle = (\rho_s, \rho_p, \rho_d)^\text{T}$ is written by
\begin{eqnarray}\label{eq:dynamical}
H'(q,t)|\psi\rangle
= {\begin{pmatrix}
H_{00}\rho_s + H_{01}\rho_p + H_{02}\rho_d \\
H_{10}\rho_s + H_{11}\rho_p + H_{12}\rho_d \\
H_{20}\rho_s + H_{21}\rho_p + H_{22}\rho_d
\end{pmatrix}}
= i\hbar{\begin{pmatrix} \dot\rho_s \\ \dot\rho_p \\ \dot\rho_d \end{pmatrix}}. \nonumber \\
\,
\end{eqnarray}
Claiming $\dot\rho_p=0$ owing to the $p$ band being negligibly populated, we obtain
$\rho_p = -(H_{10}\rho_s+H_{12}\rho_d)/H_{11}$.
Injecting this relation back into Eq.~(\ref{eq:dynamical}) yields the effective Hamiltonian as
\begin{equation}
H_{\text{eff}}(q,t)\,
= {\begin{pmatrix}
H_{00}-\frac{H_{01}H_{10}}{H_{11}} & H_{02}-\frac{H_{01}H_{12}}{H_{11}} \\
H_{20}-\frac{H_{21}H_{10}}{H_{11}} & H_{22}-\frac{H_{21}H_{12}}{H_{11}}
\end{pmatrix}}.
\end{equation}
The additional terms in the diagonal and the off-diagonal element are proportional to $\frac{F^2}{\Delta_p}$, which represent additive band energy shifts and $sd$ interband couplings, respectively, arising from the off-resonant couplings to the $p$ band.

In terms of the Pauli matrices $\bm{\sigma}=\{\sigma_x,\sigma_y,\sigma_z\}$, we obtain the modified effective Hamiltonian as
\begin{eqnarray}\label{eq:eff_H_1}
H_{\rm eff}'&&(q,t) =
\Bigg[
\bigg(\frac{\hbar\delta}{2} + 2t_{-}\cos\big(q -\theta_0\sin(\ot+\varphi)\big)\bigg) \nonumber \\
- &&\bigg(\lambda'_{-}\cos(\ot) + {F'_-}^{2}\cos(2\ot+2\varphi) + {F'_-}^{2}\bigg)
\Bigg] \sigma_z \nonumber \\
+ &&\bigg(\lambda'\cos(\omega t) + {F'}^{2}\cos(2\omega t + 2\varphi) + {F'}^{2}\bigg) \cos(\ot)\sigma_x \nonumber \\
\hlym{+} &&\bigg(\lambda'\cos(\omega t) + {F'}^{2}\cos(2\omega t + 2\varphi) + {F'}^{2}\bigg) \sin(\ot)\sigma_y \nonumber \\
\,
\end{eqnarray}
with $\delta = \omega - \omega_{sd}$.
The definitions of \hlb{$t_{-}$}, $\theta_0$, $\lambda'_{(-)}$, and $F'_{(-)}$ are listed in Table~\ref{tab:parameters}.
In the derivation of $H_{\rm eff}'$, we \hly{ignored} the trace part of the Hamiltonian, i.e., $H_{\rm eff}'=H_{\rm eff}-\frac{{\rm tr}(H_{\rm eff})}{2}\mathbb{I}$, which does not affect the topological properties of the system.

Next, we derive the approximated time-independent Hamiltonian $\Tilde{H}_{\rm eff}(q)$ for $H_{\rm eff}'(q,t)$ using the high-frequency expansion method~\cite{PhysRevA.68.013820, Eckardt_2015}.
When the Fourier series expansion of $H_{\rm eff}'(q,t)$ is given by $H_{\rm eff}'(q,t)=\Sigma_m H_m(q)e^{im\omega t}$, the second-order approximation of $\Tilde{H}_{\rm eff}(q)$ is given by
\begin{equation}
\Tilde{H}_{\rm eff}(q) = H_0 + \sum_{m>0}\frac{[H_m, H_{-m}]}{m\hbar\omega}.
\end{equation}
Neglecting the higher order terms~\cite{footnote1}, we obtain
\begin{eqnarray}\label{eq:eff_H_2}
\Tilde{H}_{\rm eff}(q) =\;
&&\left(
\frac{\hbar\delta}{2}
+2t_{-}J_0(\theta_0)\cos(q)
-{F'_-}^{2}
\right)\sigma_z \nonumber \\
&&\hlym{+}\left(
\frac{6{F'}^{2}}{\hbar\omega}t_{-}J_1(\theta_0)\sin(q)\sin(\varphi)
\hlym{+}\frac{\lambda'}{2}
\right)\sigma_x \nonumber \\
&&+\left(
\frac{6{F'}^{2}}{\hbar\omega}t_{-}J_1(\theta_0)\sin(q)\cos(\varphi)
\right)\sigma_y \nonumber \\
=\;
&&
\Big[\delta' + 2t'_{-}\cos(q)\Big]\sigma_z
+ t_v\sigma_x \nonumber \\
&&+ \, 2t_d\sin(q)\Big[\sin(\varphi)\sigma_x \hlym{+} \cos(\varphi)\sigma_y\Big],
\end{eqnarray}
where $\delta' = \hbar\delta/2 - {F'_-}^{2}$, $t'_{-} = t_{-}J_0(\theta_0)$, $t_v = \lambda'/2$, and \hly{$t_d = \frac{3{F'}^{2}}{\hbar\omega}t_{-}J_1(\theta_0)$}.
\hlb{The details of the derivation are provided in Appendix~\mbox{\ref{app:high-freq}}.}

The final expression of $\tilde{H}_{\rm eff}(q)$ in Eq.~(\ref{eq:eff_H_2}) reveals the band topology of the driven lattice system.
The terms with $t_v$ and $t_d$ correspond to the vertical and diagonal interleg links in the two-leg-ladder description [Fig.~1(a)]. 
Notably, 
\begin{eqnarray}
    t_v\, &&\propto \lambda_0 u_{sd}^{(0)} \nonumber \\
    t_d\, &&\propto \phi_0^3 \eta_{sp}^{(0)}\eta_{pd}^{(0)} (t_d^{(1)}-t_s^{(1)}),
\end{eqnarray}
indicating that the vertical links are generated by the on-site one-photon interorbital transition $|j,s\rangle \leftrightarrow |j,d\rangle$, induced by AM, while the diagonal links originate from the three-photon transitions involving site hopping, e.g., $|j,s\rangle \leftrightarrow |j,p\rangle \leftrightarrow |j,d\rangle \leftrightarrow |j+1,d\rangle$, induced by PM.

The effective Hamiltonian $\Tilde{H}_{\rm eff}(q)$ exhibits chiral symmetry at $\varphi=\pm\frac{\pi}{2}$, as $\sigma_y \Tilde{H}_{\rm eff}(q) \sigma_y = -\Tilde{H}_{\rm eff}(q)$;
this means that the spin states of the bands are restricted to the $xz$ plane \hlb{of a three-dimensional space with axes represented by Pauli matrices}, ensuring that the spin winding number across the Brillouin zone is well-defined and topologically protected by symmetry.
At $\delta'=0$, a topologically critical point emerges when $t_d=\pm t_v/2$, rendering $\Tilde{H}_{\rm eff}(q=\mp \frac{\pi}{2})=0$ \hly{for $\varphi=\frac{\pi}{2}$ and $\Tilde{H}_{\rm eff}(q=\pm \frac{\pi}{2})=0$ for $\varphi=-\frac{\pi}{2}$}.
Given the parameters of the optical lattice system at $V_0=10\,E_R$ ($E_R$ is the lattice recoil energy), as detailed in Table \ref{tab:parameters}, the ratio $|t_v/t_d|=2$ is achieved when ${\phi_0}^3/{\lambda_0} = 0.009$. This modulation condition is experimentally feasible, for example, with $\lambda_0=0.1$ and $\phi_0\approx 0.1$, which corresponds to the lattice-shaking amplitude of $0.03 a$.

Thus far, we have demonstrated that a cross-linked ladder structure can be established in a three-band optical lattice by utilizing dual-mode resonant driving. Developing an effective two-band description, we have clarified the critical role of the off-resonant $p$ band in Floquet engineering, which is essential for determining the topological characteristics of the driven lattice system. In the following section, we will confirm our theoretical findings through a direct numerical simulation of the three-band Hamiltonian $H(q,t)$ in Eq.~(\ref{eq:Bloch_Hamiltonian}).

\section{\label{sec:3}Floquet state analysis}
\subsection{\label{subsec:3a}Quasienergy spectrum}

We investigate the quasienergy spectrum of the driven three-band optical lattice system in accordance with Floquet theory~\cite{Holthaus_2016}. We numerically calculate the time-evolution operator over one driving period $T=\frac{2\pi}{\omega}$, defined as
\begin{eqnarray}
\label{eq:Evolution operator}
\hat{U}(t+T,t;q) ~&&= \mathcal{T} {\rm exp}\left[-\frac{i}{\hbar}\int_{t}^{t+T} H(q,t') dt'\right]
\end{eqnarray}
with $\mathcal{T}$ being the time-ordering operator, and obtain the quasienergy spectrum $\varepsilon_n(q)$ by directly diagonalizing $\hat{U}(t+T,t;q)$. Here, $n=0,1,2$ is the Floquet band index and $\varepsilon_n(q)\in [-\frac{\hbar\omega}{2},\frac{\hbar \omega}{2})$ is independent of the choice of time $t$.
In the calculation, we use the parameter values listed in Table~\ref{tab:parameters} and set the modulation frequency to $\omega=\omega_{sd}$.

\begin{figure}
\includegraphics{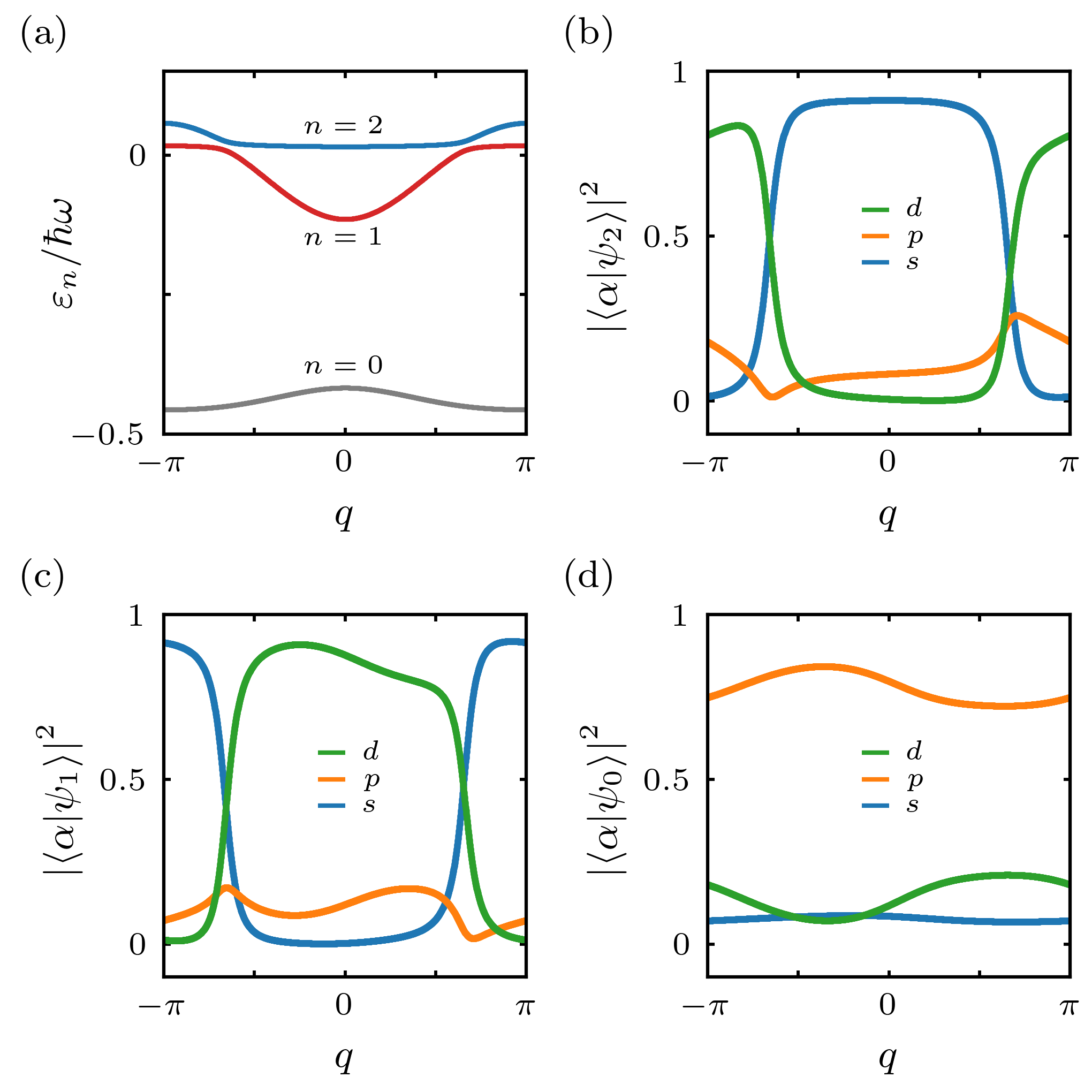}
\caption{\label{fig:quasi-energy}(a) Quasienergy spectrum $\varepsilon_n(q)$ of the three-band system driven at $\omega = \omega_{sd}$ with $\lambda_0 = 0.05$, $\phi_0 = 0.1$, and $\varphi = 0$ \hly{at $t=0$}.
The Floquet Bloch bands are indexed by $n=0,1,2$.
Fractional weights of the original orbitals $|\alpha=s,p,d\rangle$ in the (b) $n=2$, (c) $n=1$, and (d) $n=0$ Floquet bands in (a). The blue, orange, and green solid lines indicate the weights of the $s$, $p$, and $d$ orbitals, respectively.
\hly{For the case of $\varphi=\pm \pi/2$, see Appendix~\mbox{\ref{app:QS}}.}
}
\end{figure}

In Fig.~\ref{fig:quasi-energy}(a), the quasienergy spectrum is presented for $\lambda_0=0.05$, $\phi_0=0.1$, and $\varphi=0$.
The two upper ($n=1,2$) Floquet bands demonstrate the avoided crossing of the bare $s$ and $d$ bands of the stationary lattice system under the resonant driving, while the lower ($n=0$) Floquet band is located apart from the upper bands, aligned with the off-resonant $p$ band. 
In Figs.~\ref{fig:quasi-energy}(b)--\ref{fig:quasi-energy}(d), we plot the fractional weights of the $\alpha=s,p,d$ orbitals in the Floquet Bloch states $|\psi_n(q,t)\rangle$.
The Floquet Bloch states are eigenstates of $\hat{U}(t+T,t;q)$ such that
\begin{equation}\label{eq:Floquet_Bloch_state}
\hat{U}(t+T,t;q)|\psi_n(q,t)\rangle \,= e^{-i\varepsilon_n(q) T/\hbar}|\psi_n(q,t)\rangle.
\end{equation}
It is observed that the $p$ orbital contribution is minimal in the upper Floquet bands, as expected from the off-resonance nature of the $p$ band. This observation supports the validity of our use of adiabatic elimination in the previous section.

\subsection{\label{subsec:3b}Topological characteristics}

\begin{figure}[t]
\includegraphics{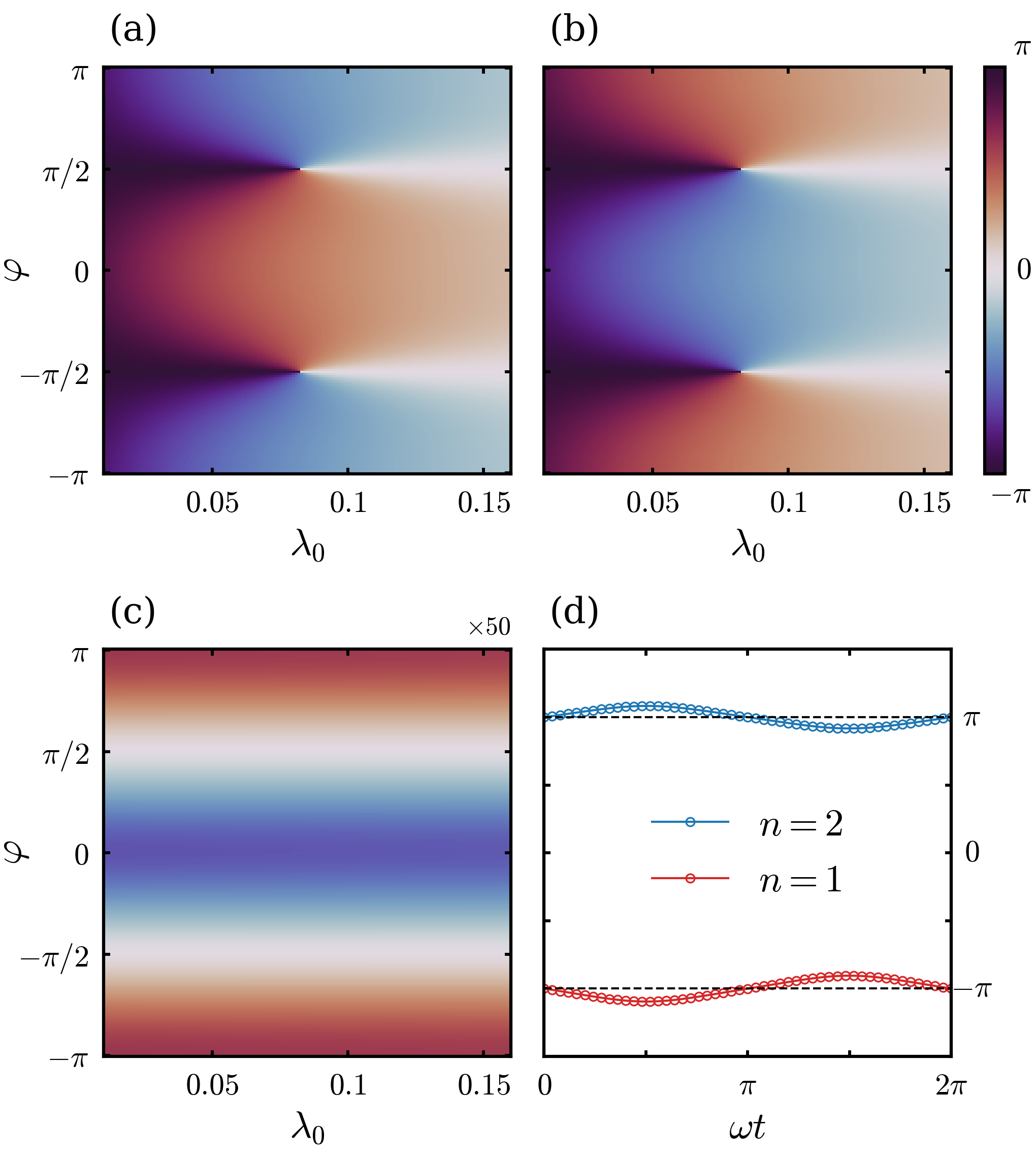}
\caption{\label{fig:Zakphase}Zak phases $\gamma_n$ of the Floquet Bloch bands at $t=0$, as a function of $\lambda_0$ and $\varphi$ for $\phi_0=0.1$: (a) $n=2$, (b) $n=1$, and (c) $n=0$. Two topological singular points are identified at $\{\varphi,\lambda_0\} =\{\pm\pi/2$, 0.082\}. In (c), the value of the Zak phase is magnified by 50 for clarity. (d) Temporal evolution of the Zak phases over one driving period, $0<t<T=\frac{2\pi}{\omega}$, for $\{\lambda_0, \varphi\}=\{0.05,\frac{\pi}{2}\}$.}
\end{figure}

To examine the topological characteristics of the driven lattice system, we calculate the Zak phases of the Floquet bands~\cite{PhysRevLett.62.2747, RevModPhys.82.3045}, which are defined over the Brillouin zone (BZ) as
\begin{equation}
\gamma_n(t) = i \int_{\text{BZ}} dq \, \langle \psi_n(q,t) | \partial_q | \psi_n(q,t) \rangle.
\end{equation}
The numerical results of $\gamma_n(t=0)$ for $\phi_0=0.1$ are illustrated in Fig.~\ref{fig:Zakphase}, as a function of the driving parameters $\lambda_0$ and $\varphi$. 
It is noted that critical points are found at $\lambda_0=0.082$ and $\varphi=\pm \frac{\pi}{2}$, accompanied by discontinuous changes in $\gamma_{1}$ and $\gamma_2$ nearby.
The effective two-band model in the previous section predicts the critical points at $\lambda_0=0.071$ for $\delta=0$ and $\phi_0=0.1$, which is in a good agreement with our numerical observations~\cite{footnote2}.
We note that when $\varphi = \pm \frac{\pi}{2}$, the Zak phase takes only the values of zero or $\pi$, while the Zak phase continuously varies in the parameter space; this is consistent with the symmetry protection condition discussed in the previous section.
Furthermore, we observe that $\gamma_1+\gamma_2 =0$ only for $\varphi=\pm\frac{\pi}{2}$, i.e., the Zak phase of the lowest $(n=0)$ Floquet band is $\gamma_0\neq 0$ for $\varphi \neq \pm \frac{\pi}{2}$ [Fig.~\ref{fig:Zakphase}(c)]; this is a characteristic of a three-band system.

In Fig.~\ref{fig:Zakphase}(d), we show the time evolution of the Zak phases for $\varphi=\frac{\pi}{2}$, revealing that they show quantized values only at $t=0$ and $\frac{T}{2}$.
For the effective two-band Floquet system, the chiral symmetry is expressed as $\sigma_y {H_{\rm eff}}'(q,t+t_0) \sigma_y = -{H_{\rm eff}}'(q,-t+t_0)$ with a proper choice of time frame $t_0$~\cite{PhysRevB.82.235114}, and we find that the symmetry condition is satisfied only with $\varphi=\pm\frac{\pi}{2}$ (mod $2\pi$) at $t_0=0$ and $\frac{T}{2}$ (mod $T$), which is consistent with the times when the Zak phases are well quantized.


As another topological characteristic of the system, we examine the entanglement entropy and spectrum~\cite{PhysRevLett.101.010504, PhysRevB.82.241102, PhysRevB.83.245132, PhysRevB.94.205422, PhysRevResearch.4.043164, PhysRevB.73.245115}.
For a 1D non-interacting fermionic system, the entanglement entropy $S$ of the many-body ground state $\vert\Psi\rangle$ is defined as
\begin{equation}
\label{eq:entanglement entropy}
S = -{\rm Tr}(\rho_{\rm A}\ln\rho_{\rm A}),
\end{equation}
where $\rho_{\rm A} = {\rm Tr}_{\rm B} |\Psi\rangle\langle\Psi|$ \hlb{is the reduced density matrix of $|\Psi\rangle$ on subsystem A.
Here, A and B denote the two subsystems that are formed by splitting the system into two equal parts.
The entanglement spectrum $\xi$ comprises the eigenvalues of the single-particle correlation matrix, $C^{lm}_{jk}= \langle\Psi| \hat{a}^{\dag}_{jl}\hat{a}_{km} |\Psi\rangle$, limited to subsystem A,
where $\hat{a}^{\dag}_{jl}\, (\hat{a}_{jl})$ denotes the creation (annihilation) operator for an atom in the Floquet Wannier state $|j,l\rangle$, localized on lattice site $j$ within subsystem A in the $n=l$ Floquet band.
The details on the calculation of $S$ and $\xi$ are provided in Appendix~\mbox{\ref{app:ES_EE}}.
}
When the system undergoes a quantum phase transition, the entanglement entropy exhibits a sharp peak~\cite{PhysRevB.73.245115} and
furthermore,
\hly{the entanglement spectrum}
unveils the system's mid-gap states. The presence of mid-gap states serves as an indication of the non-trivial topological phase of the system, which is analogous to the bulk-edge correspondence observed in edge states~\cite{PhysRevB.82.241102,PhysRevB.83.245132}, and it holds even in the case of Floquet systems~\cite{PhysRevB.94.205422,PhysRevResearch.4.043164}.

\begin{figure}
\includegraphics{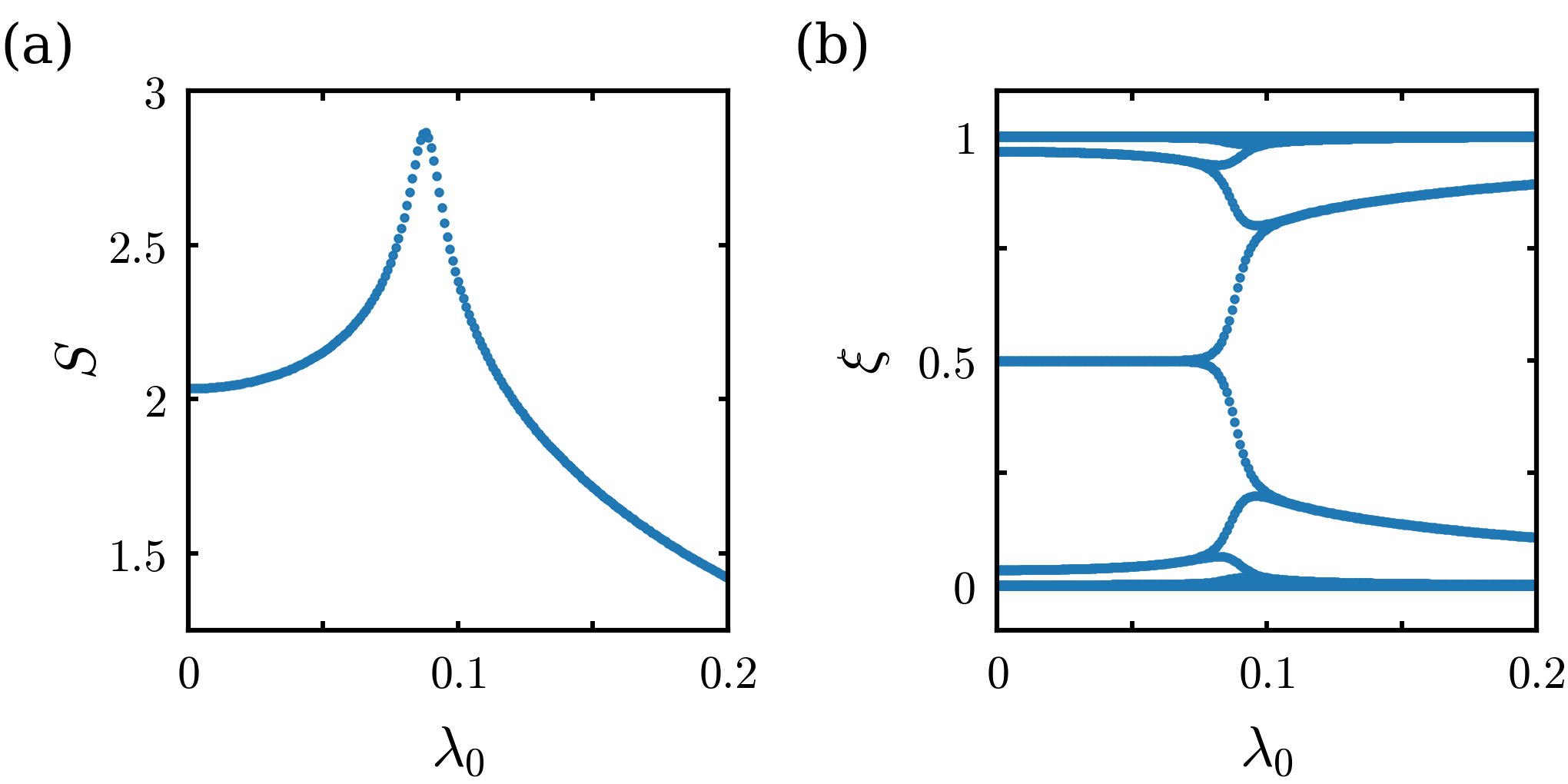}
\caption{\label{fig:Entanglement}Entanglement entropy $S$ and spectrum $\xi$ of the driven three-band system with only the $n=1$ Floquet band being filled uniformly. (a) $S$ and $\xi$ as functions of $\lambda_0$ for $\omega = \omega_{sd}$, $\varphi_0=0.1$ and $\varphi=\pi/2$. At $\lambda_0\approx 0.08$, the entanglement entropy exhibits a sharp peak, and the entanglement spectrum shows mid-gap states splitting, indicating a topological phase transition.}
\end{figure}

Figure~\ref{fig:Entanglement} presents our calculation results of the entanglement entropy and spectrum of non-interacting spinless fermions for our three-band system.
The many-body ground state $\vert\Psi\rangle$ is a uniformly filled topological Floquet band, and we choose the $n=1$ band in Fig.~\ref{fig:quasi-energy}(a) as our reference state.
When $\varphi=\pi/2$, the entanglement entropy exhibits a sharp peak at the critical point as $\lambda_0$ varies [Fig.~\ref{fig:Entanglement}(a)], indicating a topological phase transition~\cite{PhysRevB.82.241102,PhysRevResearch.4.043164}.
In the entanglement spectrum, we also observe the presence of mid-gap states and their splitting into upper and lower states at the same critical point of $\lambda_0$ [Fig.~\ref{fig:Entanglement}(b)].
These results are consistent with the Zak phase in the parameter space [Fig.~\ref{fig:Zakphase}(b)].

Finally, we remark on the edge states in our system, which are another characteristic of the topological phase~\cite{doi:10.1126/science.aaa8736, doi:10.1126/science.aaa8515}.
In our three-band system, the global bulk gap may not exist because both the $s$ and $d$ bands exhibit a similar curvature tendency, although varying in degree.
The absence of the global bulk gap implies that symmetry-protected edge states may not manifest explicitly, which was the case in our numerical investigation.

\subsection{Topological charge pumping}
\label{sec:Topological charge pumping}

When the driving parameters \{$\lambda_0, \varphi$\} vary slowly enough compared to the timescale of the driving period $T$, the system can adiabatically follow the change in driving conditions.
In other words, the long-term dynamics of the system is governed by the time-varying effective Hamiltonian, $H_{\text{eff}}(q;t) = H_{\text{eff}}(q;\{\lambda_0, \varphi\})$~\cite{PhysRevA.95.023615, WEINBERG20171}.
Using this adiabatic following, topological charge pumping can be achieved in a driven lattice system by slowly varying the driving parameters around a topological singular point, as demonstrated in recent experiments~\cite{PhysRevLett.129.053201, Walter2023}.

Given its experimental relevance, we numerically investigate the topological charge pumping effect in the driven three-band system.
A pumping protocol is considered, where the driving parameters slowly revolve around a singular point in the parameter space with the pumping cycle time $T_p$, i.e.,
\begin{eqnarray}
\label{eq:protocol}
\lambda_0(t) ~ &&= 0.1 - 0.025\cos\left({2\pi t}/{T_p}\right), \nonumber \\
\varphi(t) ~ &&= \varphi_0 + 0.5\sin\left({2\pi t}/{T_p}\right)
\end{eqnarray}
with $\varphi_0 = \pi/2$.
The system undergoes a $2\pi$ change in the Zak phase for each cycle, leading to a charge transport in which all atoms are shifted by one lattice site.
Note that this phenomenon only occurs when the trajectory of the driving parameters encircles the singular point in the parameter space, regardless of the specific details of the pumping protocol used to modulate the driving parameters~\cite{PhysRevB.27.6083, PhysRevLett.111.026802, PhysRevA.90.063638, Nakajima2016, Lohse2016}; this is why this charge pumping phenomenon is a topological one.

In the numerical simulation, the system is initially prepared in an insulating state of the Flquet band and the amount of pumped charge is calculated as $C(t) = \int_{0}^{t} dt'j(t')$, where $j(t)$ is the charge current given by $j(t) = \frac{1}{2\pi}\int_{\text{BZ}}\langle \psi(q,t)|v(q,t)|\psi(q,t)\rangle$ with velocity operator $v(q,t) = \partial H(q,t)/ \partial (\hbar q)$~\cite{PhysRevB.96.035139, RevModPhys.82.1959}.
The time evolution of the system state $|\psi (q,t)\rangle$ is calculated directly from its time-dependent Shr\"odinger equation $i \partial_t |\psi(q,t)\rangle = H(q,t)|\psi(q,t)\rangle$, including the cyclic modulations of the driving parameters.

\begin{figure}
\includegraphics{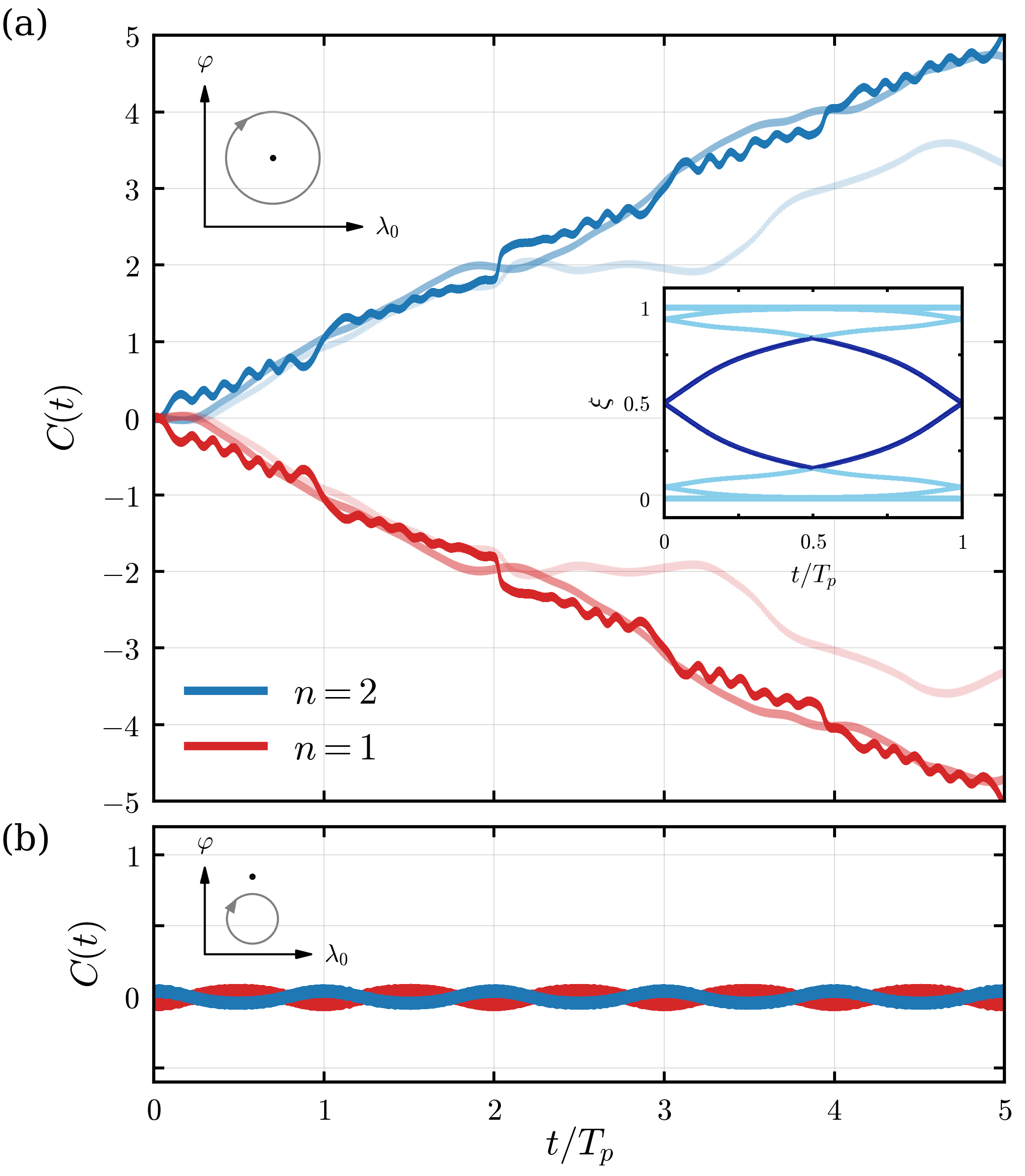}
\caption{\label{fig:pump}Numerical simulation of the topological charge pumping effect. (a) Pumped charge amount $C(t)$ as a function of the pumping time for the pumping protocol in Eq.~(\ref{eq:protocol}) with $\varphi_0 = \pi/2$ and $T_p = 900T$. The pumping protocol is sketched in the upper left inset with the dot denoting the topological singular point (Fig.~\ref{fig:Zakphase}). The solid blue and red lines indicate the results for the system initially prepared in the insulating states of the $n=2$ and $n=1$ Floquet band, respectively. The slightly faint and faintest lines show the results obtained with $T_p = 100T$ and $75T$, respectively. The inset in the middle shows the evolution of the entanglement spectrum during one pumping cycle, $T_p$. (b) Numerical results for a modified pumping protocol with $\varphi_0 = 0$ and $T_p = 900T$, where the pumping trajectory does not encircle the topological singular point in the parameter space.}
\end{figure}

In Fig.~\ref{fig:pump}(a), the pumped charge $C(t)$ is displayed as a function of time for various pumping parameter conditions.
We observe that when the change of driving parameters is slow enough, $C(t)$ increases (decreases) by unity in every pumping cycle for the $n=2$ ($n=1$) Floquet band.
The observed timescale for the adiabaticity of the charge pumping process is $T_p \approx 100T$, attributed to the local gap between the $n=1$ and $n=2$ Floquet bands, estimated as $\approx 0.01\hbar\omega$ [Fig.~\ref{fig:quasi-energy}(a)].
Furthermore, we confirm that if the trajectory of the driving parameters, such as the case of $\varphi_0 = 0$ in Eq.~(\ref{eq:protocol}), does not encircle any topological singular point in the parameter space, then the charge transport does not occur [Fig.~\ref{fig:pump}(b)].
The middle inset of Fig.~\ref{fig:pump}(a) shows the evolution of the entanglement spectrum of the driven lattice system during one pumping cycle, $T_p$.
As expected, the mid-gap states propagate like edge modes in the bulk gap~\cite{Raffaele_2000, RevModPhys.82.1959}.

\section{Symmetry in three-band model}\label{sec:4}

As predicted in the effective two-band model discussed in Sec.~\ref{subsec:Effective} and verified numerically in the preceding section, topological phases arise in the driven three-band system at $\varphi = \pm \pi/2$. Given that $\varphi = \pm \pi/2$ establishes the relationship $H(x,t)=H(-x,-t)$ in Eq.~(\ref{eq:initial Hamiltonian}), we propose that $PT$ symmetry $\mathcal{\hat{P}\hat{T}}\colon (x,t) \rightarrow (-x,-t)$ is the symmetry that protects the topological phases in this driven system. The topological phases protected by $PT$ symmetry were recently discussed in~\hlb{\mbox{\cite{PhysRevLett.118.156401, Wu2019,Ahn2018,Ahn_2019,Bouhon2019,footnote3}}}. In this section, we discuss the symmetry protection of the three-band system.

If the Floquet Hamiltonian, which is defined as $H_{F}(q,t) = i\frac{\hbar}{T}\ln[U(t+T,t;q)]$, exhibits $PT$ symmetry, it should satisfy the relation of
\begin{equation}\label{eq:PT}
U_{PT}^{\dag} {H_F(q,t+t_0)}^{\ast} U_{PT} = H_F(q,-t+t_0),    
\end{equation}
where $U_{PT}$ is a unitary matrix defined as
\begin{equation}
\label{eq:U}
U_{PT} =
\begin{pmatrix}
1 && 0 && 0 \\
0 && -1 && 0 \\
0 && 0 && 1
\end{pmatrix}
\end{equation}
for a non-interacting spinless fermionic system~\cite{RevModPhys.88.035005, PhysRevB.96.155118}.
Here, $t_0$ is the preferred time frame for the Floquet Hamiltonian \hly{$H_F(q,t)$} to exhibit $PT$ symmetry and in our system, \hly{$t_0=0$ and $\frac{T}{2}$ (mod $T$)} for $\varphi=\pm \pi/2$.
We consider the situation at \hly{$t=0$ and $t_0=0$, omitting} the time notation in the following.
On the orbital basis $|\alpha\rangle$, the Floquet state $|\psi_n(q)\rangle$ is expressed as
\begin{equation}
|\psi_n(q)\rangle
= \sum_\alpha \rho_{n\alpha} |\alpha\rangle
= \sum_\alpha |\rho_{n\alpha}|e^{i\Theta_{n\alpha}}|\alpha\rangle
\end{equation}
where $\rho_{n\alpha}$ is a complex function defined on $q$, and $\Theta_{n\alpha}$ is the argument of $\rho_{n\alpha}$.
Then, the $PT$ symmetry condition of $H_F(q)$ in Eq.~(\ref{eq:PT}) requires $U_{PT}|\psi_n(q)\rangle^{\ast} = e^{i \vartheta_n} |\psi_n(q)\rangle$, i.e., 
\begin{equation}\label{eq:PT_2}
\begin{pmatrix}
\rho_{ns}^{\ast} \\ -\rho_{np}^{\ast} \\ \rho_{nd}^{\ast}
\end{pmatrix}
= e^{i \vartheta_n}
\begin{pmatrix}
\rho_{ns} \\ \rho_{np} \\ \rho_{nd}
\end{pmatrix}
\end{equation}
with $\vartheta_n$ being a real function of $q$. This requirement can be encapsulated in two relations:
\begin{eqnarray}
    {\rm (I)}~~&& 2\Theta_{ns} = 2\Theta_{nd}~~~~~~~~\,({\rm mod}~2\pi)  \nonumber \\
    {\rm (II)}~~ && 2\Theta_{np} = 2\Theta_{ns} + \pi~~~({\rm mod}~2\pi). 
\end{eqnarray}
Here, we choose a gauge of $|\psi_n(q)\rangle$ for $\Theta_{np}$ to be $\pi/2$ and then, under this gauge fixing, $\rho_{np}$ is imaginary and $\rho_{ns}$ and $\rho_{nd}$ are real-valued.

The constraints on $|\psi_n(q)\rangle$ due to $PT$ symmetry significantly affect the Zak phase of the Floquet band.
Using Eq.~(23), the Zak phase is expressed as
\begin{equation}
\label{eq:Zak}
\gamma_n = i \int_{\text{BZ}} dq \, \langle \psi_n(q) | \partial_q | \psi_n(q) \rangle \\
= - \sum_\alpha \int_{\text{BZ}} |\rho_{n\alpha}|^2 d\Theta_{n\alpha}.
\end{equation}
This expression shows that $\gamma_n$ can be interpreted as twice the sum of the areas of the closed loops traced by $\rho_{n\alpha}$ on the complex plane.
When $PT$ symmetry is present, the enclosed area traced by $\rho_{n\alpha}$ becomes zero in general because $\rho_{np}$ is confined to the imaginary axis and $\rho_{ns} \,(\rho_{nd})$ to the real axis. Thus, the topological phase of the Floquet band is trivial with $\gamma_n=0$. However, in a special situation where $\rho_{np}$ becomes zero at $q=q_0$, the second relation in Eq.~(25) is not necessarily required so that $\rho_{ns}$ and $\rho_{nd}$ can have complex values even with the fixed gauge of $\Theta_{np}=\pi/2$; this means that as $q$ passes through $q_0$, $\rho_{ns}$ and $\rho_{nd}$ can trace paths on the complex plane and return to the real axis. In the trace, the angle between $\rho_{ns}$ and $\rho_{nd}$ must be maintained because of the first relation in Eq.~(25). Then, in the vicinity of $q=q_0$, $\Theta_{ns}$ and $\Theta_{nd}$ have identical variations of $\Delta \Theta=0$ or $\pi$ (mod $2\pi$), and it results in $\gamma_n
= - \left(|\rho_{ns}(q_0)|^2 + |\rho_{nd}(q_0)|^2\right) \Delta \Theta = 0$ or $\pi$ (mod $2\pi$),
where we use the normalization condition of $|\psi_n(q_0)\rangle$.
Consequently, the $PT$ symmetry requires the quantization of the Zak phase, thus protecting the topological phases of the three-band system.

\section{Summary}\label{sec:Summary}
We introduced a Floquet framework for controlling the topological features of a 1D optical lattice system with dual-mode resonant driving. We investigated a three-band model for the three lowest orbitals, clarifying how a cross-linked ladder forms via indirect interband coupling mediated by an off-resonant band. We provided numerical evidence for the appearance of topologically nontrivial bands in the driven system in conjunction with a phenomenon of topological charge pumping due to cyclic changes in parameters within the dual-mode resonant driving. Furthermore, we examined the role of $PT$ symmetry in protecting the band topology. The dual-mode resonant driving approach facilitates the hybridization of $s$ and $d$ orbitals with the same parity, which leads to the formation of topological bands that exhibit minimal or absent bulk gaps; this method might be used to explore the physics of topological semimetals~\cite{Burkov2016,Sun2012,Jangjan2021}. Moreover, given the unique driving mechanism relative to previous studies on shaken lattices, our dual-mode approach may provide valuable insights into the reduction of heating effects in the Floquet engineering of optical lattices~\cite{PhysRevX.4.031027, Weidinger2017, Rudner2020}.

\begin{acknowledgments}
This work was supported by the National Research
Foundation of Korea (Grants No. NRF-2023M3K5A1094811 and No. NRF-2023R1A2C3006565).
\end{acknowledgments}

\appendix
\section{Bloch Hamiltonian of the three-band system}\label{app:Bloch_H}
The Hamiltonian of a spinless single particle in a driven optical lattice takes the form of
\begin{eqnarray}
H_{\rm lat}(x,t) \,
&&= \frac{p^2}{2m} + V_{\rm lat}(x,t) \nonumber \\
&&= \frac{p^2}{2m} + \left(1+\lambda(t)\right)V_0\sin^2\left(\frac{\pi}{a}x-\phi(t)\right), \nonumber \\
\end{eqnarray}
where $p$ is the kinetic momentum of the atom, $m$ denotes its mass, $a$ is the lattice constant, and $V_0$ is the stationary lattice amplitude.
In addition, $\lambda(t)$ denotes the relative variation of the lattice amplitude, and $\phi(t)$ is the phase of the lattice potential.

The modulated lattice is generally studied in a moving frame, in which case the driving acts through an inertial force~\cite{RevModPhys.91.015005, PhysRevX.4.031027}.
When viewed from the reference frame comoving with the driven optical lattice, the Hamiltonian of the system becomes
\begin{eqnarray}
H^{(1)}_{\rm lat}(x,t)
= &&U_{x}(t)H_{\rm lat}(x,t)U^{\dag}_{x}(t)+i\hbar\dot{U}_{x}(t){U}^{\dag}_{x}(t) \nonumber \\
= &&\frac{1}{2m}\Big(p-m\dot{x}_0(t)\Big)^2 + \left(1+\lambda(t)\right)V_0\sin^2\left(\frac{\pi}{a}x\right) \nonumber \\
&&- \frac{1}{2}m\dot{x}_0(t)^2
\end{eqnarray}
by a unitary transformation with the spatial displacement operator
\begin{equation}
U_x(t) = \exp\left(\frac{ip}{\hbar}x_0(t)\right).
\end{equation}
Here, $x_0(t)$ denotes the oscillating lattice position, which is defined as $x_0(t)=\frac{a}{\pi}\phi(t)$.
To convert the time-dependent vector potential into a potential gradient, we perform an additional gauge transformation, using the time-dependent momentum displacement operator
\begin{equation}
U_p(t) = \exp\left(-\frac{ix}{\hbar}m\dot{x}_0(t)\right).
\end{equation}
Then the transformed Hamiltonian becomes
\begin{eqnarray}
H^{(2)}_{\rm lat}(x,t)
= &&U_{p}(t)H^{(1)}_{\rm lat}(x,t)U^{\dag}_{p}(t)+i\hbar\dot{U}_{p}(t){U}^{\dag}_{p}(t) \nonumber \\
= &&\frac{p^2}{2m} + \left(1+\lambda(t)\right)V_0\sin^2\left(\frac{\pi}{a}x\right)  \nonumber \\
&&\quad + m\ddot{x_0}(t)x - \frac{1}{2}m\dot{x}_0(t)^2,
\end{eqnarray}
where the last term is a global time-dependent energy shift that does not impact the system's dynamics.
Hence, by applying an appropriate unitary transformation, we can cancel it out, and the resulting Hamiltonian is
\begin{eqnarray}
H(x,t)
&&= \frac{p^2}{2m} + \left(1+\lambda(t)\right)V_0\sin^2\left(\frac{\pi}{a}x\right) + m\ddot{x_0}(t)x \nonumber \\
&&= H_0 + \lambda(t)V_{\rm stat}(x) - F(t)x,
\end{eqnarray}
where we define
\begin{eqnarray}
H_0 && =\frac{p^2}{2m} +  V_{\rm stat}(x). \\
V_{\rm stat}(x) &&= V_0\sin^2\left(\frac{\pi}{a}x\right), \\
F(t) &&= -m\ddot{x_0}(t).
\end{eqnarray}
Here, $F(t)$ represents the inertial force resulting from the phase modulation, and $V_{\rm stat}(x)$ is the stationary lattice potential.

In the tight-binding approximation, the Hamiltonian can be expressed in terms of Wannier states $|i,\alpha\rangle$ localized on lattice site $i$ in the $\alpha$ band~\cite{PhysRevX.4.031027}:
\begin{eqnarray}
H(x,t) = \sum_{ij\alpha\beta}\Big(\langle && i, \alpha|H_0|j, \beta\rangle + \lambda(t)\langle i, \alpha|V_{\rm stat}(x)|j, \beta\rangle \nonumber \\
&&- F(t)\langle i, \alpha|x|j, \beta\rangle\Big)\hat{c}^{\dag}_{i\alpha}\hat{c}_{j\beta},
\end{eqnarray}
where $\hat{c}^{\dag}_{i\alpha}\, (\hat{c}_{i\alpha})$ is the creation (annihilation) operator for the atom in the Wannier state $|i,\alpha\rangle$.
\hlg{
To restore translational symmetry, we perform a gauge transformation using the unitary operator $U^{\dag}_p(t)$.
}
Then, the Hamiltonian takes the form of
\begin{eqnarray}
H(x,t) = &&\sum_{j\alpha}\epsilon_{\alpha}\hat{c}^{\dag}_{j\alpha}\hat{c}_{j\alpha}
-\sum_{jl\alpha}t^{(l)}_{\alpha}e^{-il\theta(t)}\hat{c}^{\dag}_{j\alpha}\hat{c}_{j+l\:\alpha} \nonumber \\
&&+\sum_{jl\alpha\beta}\Big(\lambda(t)u^{(l)}_{\alpha\beta}-F(t)\eta^{(l)}_{\alpha\beta}\Big)e^{-il\theta(t)}\hat{c}^{\dag}_{j\alpha}\hat{c}_{j+l\:\beta}, \nonumber \\
\end{eqnarray}
where we introduce the following parameters:
\begin{eqnarray}
\epsilon_\alpha &&= \langle j,\alpha |H_0| j,\alpha\rangle, \\
t^{(l)}_\alpha &&=
\hlgm{
\begin{cases}
-\langle j,\alpha |H_0| j+l,\alpha\rangle &(l \neq 0) \\
0 &(l=0)
\end{cases}
}
,\\
u^{(l)}_{\alpha\beta} &&= \langle j,\alpha|V_{\rm stat}(x)|j+l,\beta\rangle, \\
\eta^{(l)}_{\alpha\beta} &&=
\hlgm{
\begin{cases}
\langle j,\alpha|x|j+l,\beta\rangle &(\alpha \neq \beta) \\
0 &(\alpha=\beta)
\end{cases}
}
,\\
\theta(t) &&= -\frac{a}{\hbar} \int_{0}^{t} dt' \,F(t').
\end{eqnarray}
Here, $\epsilon_\alpha$ represents the on-site energy, and $t^{(l)}_\alpha$ denotes the hopping amplitude between the Wannier states in the $\alpha$ band separated by $l$ lattice sites.
In addition, $u^{(l)}_{\alpha\beta}$ and $\eta^{(l)}_{\alpha\beta}$ correspond to the lattice potential and lattice displacement matrix elements for interorbital transitions separated by $l$ lattice sites, respectively.
Lastly, $\theta(t)$ represents the time-dependent Peierls phase~\cite{PhysRevA.92.043621}.

By Fourier transforming this tight-binding model Hamiltonian, we obtain the Bloch Hamiltonian for quasimomentum $q$ in the presence of amplitude and phase modulations as follows:
\begin{eqnarray}
H(q,t) = \sum_{\alpha}\Big(\epsilon_{\alpha}-\sum_{l>0}2t^{(l)}_{\alpha}{\cos}[l(q-\theta(t))]\Big)\hat{c}^{\dag}_{q\alpha}\hat{c}_{q\alpha} \nonumber \\
+\sum_{l\alpha\beta}\Big(\lambda(t)u^{(l)}_{\alpha\beta}-F(t)\eta^{(l)}_{\alpha\beta}\Big)e^{il(q-\theta(t))}\hat{c}^{\dag}_{q\alpha}\hat{c}_{q\beta}. \nonumber \\
\end{eqnarray}
Here, $q$ is expressed in units of $1/a$.
In this work, we consider a model system that includes only the three lowest bands, indexed by $\alpha \in \{s, p, d\}$. Considering the lowest-order effects of lattice modulation, the Bloch Hamiltonian of the three-band system is given by
\begin{equation}
H(q,t)
= \begin{pmatrix}
\epsilon_{s}'(q,t) && -F(t)\eta^{(0)}_{sp} && \lambda(t)u^{(0)}_{sd} \\
-F(t)\eta^{(0)}_{ps} && \epsilon_{p}'(q,t) && -F(t)\eta^{(0)}_{pd} \\
\lambda(t)u^{(0)}_{ds} && -F(t)\eta^{(0)}_{dp} && \epsilon_{d}'(q,t)
\end{pmatrix},
\end{equation}
where $\epsilon_{\alpha}'(q,t) = \epsilon_{\alpha}-2t^{(1)}_{\alpha}\cos(q-\theta(t))+\lambda(t)u^{(0)}_{\alpha\alpha}$.

\section{\hlys{High-frequency expansion method}}\label{app:high-freq}
In Floquet theory, the high-frequency expansion method is one of the useful techniques for analyzing periodically driven $(\omega)$ quantum systems. The main idea of the high-frequency expansion method is to separate the effects of periodic driving on the system into fast and slow parts. By using perturbation theory (in powers of $\omega^{-1}$), it converts the motion for the slow part of the system into a time-independent effective Hamiltonian, making it easier to analyze the system's dynamics. This approach is valid when the driving frequency is sufficiently larger than any other relevant energy scale of the system.

From Eq.~\ref{eq:eff_H_1}, one can obtain the coefficients $H_m(q)$ of the Fourier series expansion of $H_{\rm eff}'(q,t)=\Sigma_m H_m(q)e^{im\omega t}$, which are given by
\begin{eqnarray}
&&H_0 = \left(\frac{\hbar\delta}{2} + 2t_{-}J_0(\theta_0)\cos(q) - {F'_-}^{2}\right)\sigma_z
+ \frac{\lambda'}{2}\sigma_x, \nonumber \\
&&H_1 = -\left(2it_{-}J_1(\theta_0)e^{i\varphi}\sin(q) + {\lambda'_-}^{2}\right)\sigma_z
+ \frac{{F'}^2}{2}e^{i2\varphi}\hlym{\sigma_{+}} \nonumber \\
&&\qquad \quad + {F'}^2\hlym{\sigma_{-}}, \nonumber \\
&&H_2 = \left(2t_{-}J_2(\theta_0)e^{i2\varphi}\cos(q) - \frac{{F'_-}^{2}}{2}e^{i2\varphi}\right)\sigma_z
+ \frac{\lambda'}{2}\hlym{\sigma_{-}}, \nonumber \\
&&H_3 = -2it_{-}\sin(q)e^{i3\varphi}J_3(\theta_0)\sigma_z
+ \frac{F'^2}{2}e^{i2\varphi}\hlym{\sigma_{-}}, \nonumber \\
&&H_{m=\rm even} = 2t_{-}\cos(q)e^{im\varphi}J_m(\theta_0)\hlym{\sigma_z}, \nonumber \\
&&H_{m=\rm odd} = -2it_{-}\sin(q)e^{im\varphi}J_m(\theta_0)\hlym{\sigma_z}, \nonumber \\
&&H_{-m} = H^\dag_m
\end{eqnarray}
with $\sigma_\pm = (\sigma_x \pm i\sigma_y)/2$ and $J_m$ being the $m$th order Bessel function of the first kind.
Here, $m=\rm even(odd)$ denotes the even (odd) integers greater than 3.

Using the high-frequency expansion method~\cite{PhysRevA.68.013820, Eckardt_2015}, the time-independent effective Hamiltonian $\Tilde{H}_{\rm eff}(q)$ can be perturbatively obtained as $\Tilde{H}_{\rm eff}=\sum_{k=0}^{\infty}\mathcal{H}^{(k)}(\frac{1}{\hbar\omega})^k$.
The coefficients for the leading terms are provided by
\begin{eqnarray}
\mathcal{H}^{(0)} = &&H_0, \nonumber \\
\mathcal{H}^{(1)} = &&\sum_{m \neq 0}\frac{H_m H_{-m}}{m}, \nonumber \\
\mathcal{H}^{(2)} = &&\sum_{m \neq 0}\Bigg(\frac{[H_{-m}, [H_0,H_m]}{2m^2} \nonumber \\
&& \quad + \sum_{m' \neq 0,m}\frac{[H_{-m'},[H_{m'-m},H_m]]}{3mm'}\Bigg).
\end{eqnarray}
Then the effective Hamiltonian truncated to the first-order term $\mathcal{H}^{(1)}$ is given as
\begin{eqnarray}
\Tilde{H}_{\rm eff}(q)
\approx\;&& \mathcal{H}^{(0)} + \mathcal{H}^{(1)}\left(\frac{1}{\hbar\omega}\right) \nonumber \\
=\;&& H_0 + \sum_{m>0}\frac{[H_m, H_{-m}]}{m\hbar\omega} \nonumber \\
\approx\;&& \left(\frac{\hbar\delta}{2} + 2t_{-}J_0(\theta_0)\cos(q) - {F'_-}^{2}\right)\sigma_z \nonumber \\
&& \hlym{+} \left(\frac{6{F'}^{2}}{\hbar\omega}t_{-}J_1(\theta_0)\sin(q)\sin(\varphi) \hlym{+} \frac{\lambda'}{2}\right)\sigma_x \nonumber \\
&& + \left(\frac{6{F'}^{2}}{\hbar\omega}t_{-}J_1(\theta_0)\sin(q)\cos(\varphi)\right)\sigma_y \nonumber \\
=\;&& \Big[\delta' + 2t'_{-}\cos(q)\Big]\sigma_z + t_v\sigma_x \nonumber \\
&&+ \, 2t_d\sin(q)\Big[\sin(\varphi)\sigma_x \hlym{+} \cos(\varphi)\sigma_y\Big],
\end{eqnarray}
where $\delta' = \hbar\delta/2 - {F'_-}^{2}$, $t'_{-} = t_{-}J_0(\theta_0)$, $t_v = \lambda'/2$, and \hly{$t_d = \frac{3{F'}^{2}}{\hbar\omega}t_{-}J_1(\theta_0)$}.
Here, we ignored the terms involving the second-order Bessel function $J_2(\theta_0)$ and the higher-order terms of $\lambda_0$ and $F_0^2$, as they are negligible compared to the other terms, given our parameter values in Table~\ref{tab:parameters}.

\begin{figure*}[t!]
\includegraphics{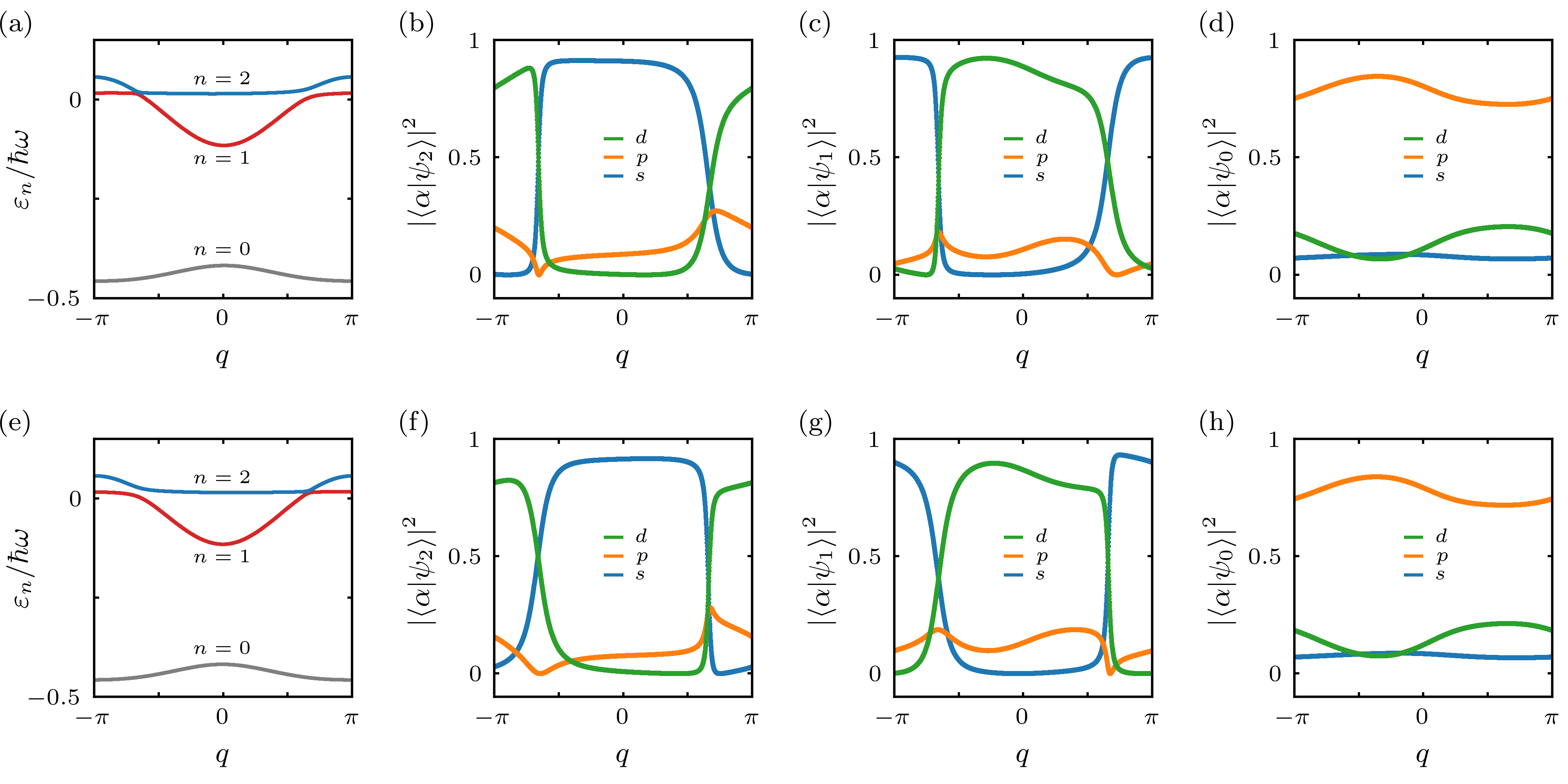}
\caption{\label{fig:pm0.5pi}Quasienergy spectrum $\varepsilon_n(q)$ of the three-band system driven at $\{\omega,\lambda_0,\phi_0\} = \{\omega_{sd},0.05,0.1\}$ for (a) $\varphi=\pi/2$ and (e) $\varphi=-\pi/2$ \hly{at $t=0$}.
The Floquet Bloch bands are indexed by $n=0,1,2$.
Fractional weights of the original orbitals $|\alpha=s,p,d\rangle$ in the (b) $n=2$, (c) $n=1$, and (d) $n=0$ Floquet bands for $\varphi=\pi/2$, and (f) $n=2$, (g) $n=1$, and (h) $n=0$ Floquet bands for $\varphi=-\pi/2$.
The blue, orange, and green solid lines indicate the weights of the $s$, $p$, and $d$ orbitals, respectively.}
\end{figure*}

\section{Quasienergy spectrum for $\varphi=\pm\pi/2$}\label{app:QS}
In Fig.~\ref{fig:quasi-energy}, we present the quasienergy spectrum and the fractional weights of the original orbitals in the Floquet Bloch states for $\lambda_0=0.05$, $\phi_0=0.1$, and $\varphi=0$.
In this section, we also examine the cases for other values of $\varphi$, specifically $\varphi=\pm\pi/2$ [Fig.~\ref{fig:pm0.5pi}].

From Eq.~\ref{eq:Bloch_Hamiltonian} and \ref{eq:modulations}, one can observe that when the relative phase $\varphi$ changes from $\pi/2$ to $-\pi/2$, the Bloch Hamiltonian of our three-band system also changes from $H(q,t)$ to $H(-q,-t)$.
Since the quasienergy does not depend on time [Eq.~\ref{eq:Floquet_Bloch_state}], $H(-q,-t)$ should exhibit an inverted quasienergy spectrum with respect to $q$ compared to $H(q,t)$ [Fig.~\ref{fig:pm0.5pi}(a) and (e)].

\section{\hlys{Entanglement spectrum and entropy}}\label{app:ES_EE}
As mentioned in Sec.~\ref{subsec:3b}, the single-particle entanglement spectrum $\xi$ is defined as the set of eigenvalues of the correlation matrix~\cite{PhysRevB.83.245132}, which is given by
\begin{equation}
C^{lm}_{jk}
= \langle\Psi| \hat{a}^{\dag}_{jl}\hat{a}_{km} |\Psi\rangle
= \langle\hat{a}^{\dag}_{jl}\hat{a}_{km}\rangle.
\end{equation}
Here, $\hat{a}^{\dag}_{jl}\, (\hat{a}_{jl})$ denotes the creation (annihilation) operator for an atom in the Floquet Wannier state $|j,l\rangle$, localized at lattice site $j$ within subsystem A (one of the two halves of the original system) in the $n=l$ Floquet band.
After applying a Fourier transform, the correlation matrix can be expressed in terms of Floquet Bloch states as follows:
\begin{eqnarray}
C^{lm}_{jk}
&&= \sum_{q} e^{iq(j-k)}\langle \hat{a}^{\dag}_{ql}\hat{a}_{qm}\rangle \nonumber \\
&&= \sum_{q} e^{iq(j-k)} \sum_{\alpha} \langle\psi_l(q)|\alpha\rangle \langle\alpha|\psi_m(q)\rangle.
\end{eqnarray}
In this expression, $\langle\alpha|\psi_m(q)\rangle$ represents the coefficient of the $n=m$ Floquet state in the original Wannier basis, where $\alpha \in \{s, p, d\}$.
By calculating the Floquet Bloch states, which are the eigenstates of the one-cycle time-evolution operator [Eq.~(\ref{eq:Floquet_Bloch_state})], we can obtain all the coefficients $\langle\alpha|\psi_m(q)\rangle$.
These coefficients are then used to construct the correlation matrix.
Diagonalizing this matrix yields the entanglement spectrum.

Meanwhile, due to the Wick's theorem, there is a special relation between the correlation matrix and the reduced density matrix, given by~\cite{PhysRevResearch.4.043164}
\begin{equation}\label{eq:Xi}
\Xi_j = \ln({\xi_j}^{-1}-1), \qquad j \in {\rm A}.
\end{equation}
Here, $\xi_j$ are the eigenvalues of the correlation matrix (i.e., the entanglement spectrum) and $\Xi_j$ are the eigenvalues of the entanglement Hamiltonian $H_{\rm A}$, defined as
\begin{equation}
\rho_{\rm A} = \frac{1}{Z}e^{-H_{\rm A}},
\end{equation}
where $\rho_{\rm A}$ is the reduced density matrix, and $Z={\rm Tr}(e^{-H_{\rm A}})=\displaystyle\prod_{j}(1+e^{-\Xi_j})$.
Using Eq.~(\ref{eq:Xi}), the entanglement entropy $S$ can be expressed in terms of the entanglement spectrum $\xi$ as follows~\cite{PhysRevResearch.4.043164}:
\begin{eqnarray}
S
&&= -{\rm Tr}(\rho_{\rm A}\ln\rho_{\rm A}) \nonumber \\
&&= -{\rm Tr}\left[\frac{1}{Z}e^{-H_{\rm A}}\ln\left(\frac{1}{Z}e^{-H_{\rm A}}\right)\right] \nonumber \\
&&= \sum_j \ln(1+e^{-\Xi_j}) + \frac{1}{Z}{\rm Tr}(H_A e^{-H_A}) \nonumber \\
&&= \sum_j \left[\ln(1+e^{-\Xi_j}) + \frac{\Xi_j}{e^{\Xi_j}+1} \right] \nonumber \\
&&= -\sum_j \left[\xi_j\ln\xi_j + (1-\xi_j)\ln(1-\xi_j)\right].
\end{eqnarray}
In this study, we obtained the entanglement spectrum from the correlation matrix, and then calculated the entanglement entropy from the entanglement spectrum.



\begin{thebibliography}{99}
\bibitem{footnote1}
We ignore the terms with the second order Bessel function $J_2(\theta_0)$ and the higher order terms of $\lambda_0$ and ${F_0}^2$.
\bibitem{footnote2}
Different from the case when $\delta'=0$, where the gap closes at $q=\pm 0.5\pi$, in this instance, the gap closing occurs at $q=\pm 0.66\pi$ due to the presence of the energy shift ${F'_{-}}^2$ which results in $\delta'\neq0$.
\end{thebibliography}
\end{document}